\documentclass[%
 reprint,
superscriptaddress,
 amsmath,amssymb,
 aps,
prc,
]{revtex4-2}

\usepackage{graphicx}
\graphicspath{{fig/}}
\usepackage{dcolumn}
\usepackage{bm}
\usepackage{braket}
\usepackage[bookmarks=false]{hyperref}
  \hypersetup{pdfpagemode={UseOutlines},
  bookmarksopen=true,
  bookmarksopenlevel=0,
  hypertexnames=false,
  colorlinks=true,
  citecolor=blue,
  linkcolor=blue,
  urlcolor=blue,
  allcolors=blue,
  pdfstartview={FitV},
  unicode,
  breaklinks=true,
  }

\usepackage{color}  

\usepackage[normalem]{ulem}

\begin{document}

\title{Significance of the refraction effect on the $p$-$d$ elementary process in the ($p$,$pd$) reaction}

  \author{Kazuki Yoshida}
  \email[]{yoshida.kazuki@jaea.go.jp}
  \affiliation{Advanced Science Research Center, Japan Atomic Energy Agency, Tokai, Ibaraki 319-1195, Japan}

  \author{Yoshiki Chazono}
  \affiliation{RIKEN Nishina Center for Accelerator-Based Science, 2-1 Hirosawa, Wako 351-0198, Japan}

  \author{Kazuyuki Ogata}
  \affiliation{Department of Physics, Kyushu University, Fukuoka 819-0395, Japan}
  \affiliation{Research Center for Nuclear Physics, Ibaraki, Osaka 567-0047, Japan}

\date{\today}
\begin{abstract}
\begin{description}
  \item[Background]
  The proton-induced deuteron knockout reaction, ($p$,$pd$), is one of the interests in studies probing the deuteron-like $p$-$n$ correlation in nuclei.
  According to a recent study of the inclusive deuteron-induced reaction, $(d,d'x)$, the refraction effect of the deuteron has a significant effect on the elementary process, nucleon-deuteron ($N$-$d$) binary scattering inside a nucleus, of the reaction.
  In the paper, it is shown that proper treatment of the local $N$-$d$ relative momentum in the elementary process is crucial in $(d,d'x)$ reactions at $100$~MeV and below. 
  \item[Purpose]
  In the present work, we investigate the deuteron refraction effect in the exclusive ($p$,$pd$) reactions.
  We also discuss the incident energy dependence of the refraction effect.
  \item[Method]
  The refraction effect on the $p$-$d$ elementary process is taken into account by the local semiclassical approximation to the distorted waves.
  The results are compared with those obtained with the asymptotic momentum approximation, which is standardly applied to the distorted wave impulse approximation framework.
  \item[Result]
  It is shown that the refraction effect drastically changes the energy sharing distribution of the $^{16}$O($p$,$pd$)$^{14}$N reaction at 101.3~MeV and gives a better agreement with experimental data.
  In contrast, it is confirmed that the effect is negligibly small at 250~MeV.  
  \item[Conclusion]
  We have clarified that the deuteron refraction effect is significant in the $^{16}$O($p$,$pd$)$^{14}$N reaction at 101.3~MeV and the experimental data are well reproduced.
  The refraction effect plays a significant role in both the shape and magnitude of the ($p$,$pd$) cross section, while the effect is negligible at 250~MeV.
\end{description}
\end{abstract}

\maketitle

\section{Introduction}
  Quasifree knockout reactions have been used for more than half a century to probe both the single-particle (s.p.) nature of nucleons~\cite{Jacob66,Jacob73,Chant77,Chant83,Kitching85,Cowley91,Wakasa17,Noro20,Noro23} and the $\alpha$ cluster states~\cite{Gottschalk70,Bachelier73,Bachelier76,Roos77,Landaud78,Nadasen80,Carey81,Carey84,Wang85,Nadasen89,Yoshimura98,Neveling08,Cowley09,Mabiala09,Typel14,Yoshida16,Lyu18,Yoshida18,Yoshida19,Lyu19,Cowley21,Taniguchi_48Ti,Tanaka21,Yoshida22_Po,Edagawa23} of nuclei.
  These reactions have been described by the distorted wave impulse approximation (DWIA) framework~\cite{Chant77,Chant83} and the experimental data have been well understood within this framework.

  Recently, more exotic clusters, e.g., $d$, $t$, and $^{3}$He, have become of interest~\cite{Typel13,Zhang17}.
  These clusters as well as the traditional $\alpha$ cluster will be intensively studied by the ONOKORO project~\cite{onokoro_en} using cluster knockout reactions mainly at HIMAC (Heavy Ion Medical Accelerator in Chiba), RCNP (Research Center for Nuclear Physics, Osaka University), and RIKEN RIBF (Radioactive Isotope Beam Factory) in Japan.

  The proton-induced deuteron knockout reaction, $(p,pd)$, is expected to be a probe for the $p$-$n$ spatial and spin-isospin correlations, or the existence of a deuteron, inside a nucleus.
  One of the difficulties in describing the $(p,pd)$ reaction is the fragileness of the deuteron, which is not explicitly included in the standard DWIA.
  In order to incorporate this nature of the deuteron into a reaction theory, the continuum-discretized coupled-channels (CDCC) wave function~\cite{Kamimura86,Austern87,Yahiro12} was recently implemented into the DWIA framework, and the breakup and the reformation process of the $p$-$n$ pair in the ($p$,$pd$) reaction are studied within a new framework called CDCCIA~\cite{Chazono22}.
  Another difficulty is that the refraction of the deuteron by the nuclear potential is larger compared to that of the nucleon.
  According to a recent study on inclusive ($d$,$d'x$) reactions~\cite{Nakada23} using the semiclassical distorted wave model~\cite{Luo91,Watanabe99}, the refraction effect on the $p$-$d$ elementary process has a significant impact.
  The same effect is expected in exclusive $(p,pd)$ reactions and it has to be also considered in combination with the CDCCIA framework.

  It should be noted that, in Ref.~\cite{Yoshida16}, the refraction effect in the $^{120}$Sn$(p,p\alpha)^{116}$Cd was investigated and found to be small.
  The main reason for this result is the strong absorption of $\alpha$ in the region where the refraction is important.
  In other words, both the real and imaginary potentials for $\alpha$ are strong.
  Thus, the significant effect of the deuteron refraction found in Ref.~\cite{Nakada23} will suggest that the refraction of the deuteron is much more important than its absorption.

  In the present work, we focus on the deuteron refraction effect on the $p$-$d$ elementary process in the ($p$,$pd$) reaction within the DWIA framework and discuss to what extent the $(p,pd)$ cross section is affected.
  There is a limited number of ($p$,$pd$) reaction experiments~\cite{Kitching75,Ero81,Samanta82,Warner85,Warner86,Terashima18}.
  In Ref.~\cite{Joshi16}, the data of $^{16}$O($p$,$pd$)$^{14}$N$^*_{3.95~\mathrm{MeV}}$ at 101.3~MeV~\cite{Samanta82} are reanalyzed using the finite-range DWIA (FR-DWIA) formalism with two types of $p$-$d$ interactions (``all attractive'' and ``repulsive core'').

  This paper is organized as follows.
  In Sec.~\ref{sec:framework} we recapitulate the DWIA framework with the local semiclassical momentum approximation (LSCA)~\cite{Luo91, Watanabe99} and the asymptotic momentum approximation (AMA)~\cite{Yoshida16}.
  We discuss the finite-range nature of the $p$-$d$ effective interaction being partially taken into account in the position dependence of its matrix element in DWIA with LSCA.
  In Sec.~\ref{sec:result} the inputs for the DWIA calculations are explained and the effect of the local momentum on the ($p$,$pd$) cross section is discussed based on DWIA calculations with LSCA and AMA.
  The relation between the present result and the FR-DWIA calculation~\cite{Joshi16} is also discussed in Sec.~\ref{sec:refraction}.
  Finally a summary is given in Sec.~\ref{sec:summary}.

\section{Theoretical framework}
  \label{sec:framework}
  In the present work, we basically follow the DWIA framework~\cite{Chant77,Chant83,Wakasa17,Ogata23} to describe the ($p$,$pd$) reaction.
  Practical calculations are performed using a newly published code \textsc{pikoe}~\cite{Ogata23}.
  The incoming proton, the outgoing proton, and the outgoing deuteron are denoted as particles 0, 1, and 2, respectively, and the target (residual) nucleus is denoted by A (B).
  The total energy and the wave number vector are represented by $E_i$ and $\bm{K}_i$ $(i = 0, 1, 2, \mathrm{A, and~B})$, respectively. 
  All the quantities with the superscript L are evaluated in the laboratory (target rest) frame of the ($p$,$pd$) reaction, while those without the superscript are evaluated in the center-of-mass frame.
  As a characteristic ingredient in the present study, we use the local semiclassical approximation (LSCA)~\cite{Luo91,Watanabe99} to describe the propagation of the distorted waves under nuclear potentials, which is nothing but the refraction of the reaction particle.
  This method, DWIA with LSCA, has already been applied to ($p$,$p\alpha$) analyses in Ref.~\cite{Yoshida16}.

\begin{figure}[htpb]
    \centering
    \includegraphics[width=0.40\textwidth]{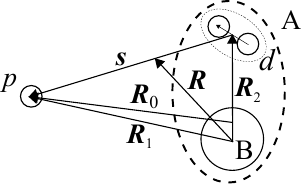}
    \caption{Definition of the coordinates of the A($p$,$pd$)B reaction.}
    \label{fig:coordinate}
  \end{figure}

  Neglecting the spin degrees of freedom in the scattering waves and the elementary process, 
  the DWIA $T$ matrix of the $(p,pd)$ reaction is given by~\cite{Chant77,Wakasa17,Chazono21,Ogata23}
  \begin{align}
    T
    &=
    \int d\bm{R}\, d\bm{s}\, \chi_{1}^{*(-)}(\bm{R}_1) \chi_{2}^{*(-)}(\bm{R}_2)  t_{pd}(\bm{s})  \chi_{0}^{(+)}(\bm{R}_0) \varphi_{d}(\bm{R}_2).
    \label{eq:t-matrix}
  \end{align}
  The deuteron cluster is assumed to be bound in $s$ wave, as we discuss such experimental data~\cite{Samanta82} in this study.
  The coordinates are defined as shown in Fig.~\ref{fig:coordinate}, and $\chi_i(\bm{R}_i)$ is the distorted wave describing the scattering state of particle $i$ with respect to the coordinate $\bm{R}_i$.
  The distorted waves with the superscripts $(+)$ and $(-)$ satisfy the outgoing and incoming boundary conditions, respectively.
  $\varphi_{d}$ is a normalized bound state wave function of the deuteron and B, and its spectroscopic factor $S_d$ is explicitly considered in Eqs.~(\ref{eq:tdx-lsca}) and (\ref{eq:tdx-ama}) below.
  $t_{pd}$ is the $p$-$d$ effective interaction which describes the $p$-$d$ elastic scattering.
  In this study, it is assumed that $t_{pd}$ does not change the $p$-$n$ relative state and the $p$-$n$ pair remains in the deuteron ground state and its breakup and reformation are not considered.

  LSCA~\cite{Luo91,Watanabe99} assumes that the short-range propagation from a reference point $\bm{R}$ to $\bm{R}+\Delta\bm{R}$ is described by a plane wave with a local momentum $\bm{K}(\bm{R})$ at the reference point $\bm{R}$:
  \begin{align}
    \chi_i(\bm{R}+\Delta\bm{R})
    \approx
    \chi_i(\bm{R})e^{i\bm{K}_i(\bm{R})\cdot\Delta\bm{R}}.
    \label{eq:lsca}
  \end{align}
  This is the first-order approximation of a translation by $\Delta \bm{R}$ of $\chi_i(\bm{R})$. 
  The norm of $\bm{K}_i(\bm{R})$ is determined by local energy conservation and the direction of $\bm{K}_i(\bm{R})$ is taken to be parallel with the flux of $\chi_i(\bm{R})$.
  LSCA has been applied to the scattering of nucleons~\cite{Luo91,Watanabe99,Minomo10}, $\alpha$ particles~\cite{Yoshida16}, and deuterons~\cite{Nakada23}.
  Applying LSCA to Eq.~(\ref{eq:t-matrix}), one obtains the DWIA $T$ matrix with LSCA~\cite{Yoshida16} as
  \begin{align}
    &T^{\mathrm{LSCA}}
    =
    \int d\bm{R}\, 
    F(\bm{R})\varphi_{d}(\bm{R}) \tilde{t}_{pd}(\bm{\kappa}(\bm{R}),\bm{\kappa}'(\bm{R})),
    \label{eq:T-lsca} \\
    &F(\bm{R})
    \equiv 
    \chi_{1}^{*(-)}(\bm{R})
    \chi_{2}^{*(-)}(\bm{R})
    \chi_{0}^{(+)}(\bm{R})
    e^{-i\bm{K}_0(\bm{R})\cdot\bm{R}A_d/A}, \\
    &\tilde{t}_{pd}(\bm{\kappa}(\bm{R}),\bm{\kappa}'(\bm{R}))
    \equiv
    \int d\bm{s}\,
    e^{-i\bm{\kappa}'(\bm{R})\cdot \bm{s}}
    t_{pd}(\bm{s})
    e^{i\bm{\kappa}(\bm{R})\cdot \bm{s}},
    \label{eq:small-t}
  \end{align}
  where $\bm{\kappa}(\bm{R})$ and $\bm{\kappa}'(\bm{R})$ are the local $p$-$d$ relative momenta in the initial and final states, respectively, which are given by
  \begin{align}
    \bm{\kappa}(\bm{R}) &\equiv
    \frac{A+1}{A}\frac{A_d}{A_d+1} \bm{K}_0(\bm{R}) - \frac{1}{A_d+1} \bm{K}_d(\bm{R}), \\
    \bm{\kappa}'(\bm{R}) &\equiv
    \frac{A_d}{A_d+1} \bm{K}_1(\bm{R}) - \frac{1}{A_d+1} \bm{K}_2(\bm{R})
    \label{eq:local-relative-mom}
  \end{align}
  as discussed in Ref.~\cite{Yoshida16}.
  $A_d$ and $A$ are the mass numbers of the deuteron and the target nucleus, respectively.
  $\bm{K}_d$ is determined to satisfy the momentum conservation of the $p$-$d$ system:
  \begin{align}
    \bm{K}_d(\bm{R}) =
    \bm{K}_1(\bm{R}) + \bm{K}_2(\bm{R}) - \frac{A+1}{A} \bm{K}_0(\bm{R}).
  \end{align}
  In the numerical calculation of Eq.~(\ref{eq:T-lsca}), in many cases, $\tilde{t}_{pd}$ parametrized in momentum space has been used, rather than calculating the right-hand side of Eq.~(\ref{eq:small-t}).
  In other words, the integration over $\bm{s}$ seems to be unnecessary in the numerical calculation.
  Because of this, sometimes Eq.~(\ref{eq:T-lsca}) is referred to as the zero-range approximation.
  However, it is obvious from Eq.~(\ref{eq:small-t}) that the finite-range nature of $t_{pd}(\bm{s})$ is explicitly taken into account.
  Furthermore, in DWIA with LSCA, the arguments of $\tilde{t}_{pd}$ depend on $\bm{R}$, which is not the case with DWIA with AMA as one sees from Eqs.~(\ref{eq:tdx-ama}) and (\ref{eq:tbar-ama}) below.
  The $\bm{R}$ dependence of $\tilde{t}_{pd}$ can be understood that the original six-dimensional nature is more respected than in DWIA with AMA.
  We will return to this point later.

  In the calculation of the $(p,pd)$ cross section, $\left|T^{\mathrm{LSCA}}\right|^2$ is obviously given by
  \begin{align} 
    \left|T^{\mathrm{LSCA}}\right|^2
    &=
    \int d\bm{R} \,
    \tilde{t}_{pd}(\bm{\kappa}(\bm{R}),\bm{\kappa}'(\bm{R}))
    F(\bm{R})
    \varphi_{d}(\bm{R})
    \nonumber \\
    &\times
    \int d\bm{R}'\,
    \tilde{t}_{pd}^{*}(\bm{\kappa}(\bm{R}'),\bm{\kappa}'(\bm{R}'))
    F^*(\bm{R}')
    \varphi_{d}^*(\bm{R}').
    \label{eq:app-t-lsca-square}
  \end{align}
  We make an additional approximation in Eq.~(\ref{eq:app-t-lsca-square}) as
  \begin{align} 
    &\tilde{t}_{pd}(\bm{\kappa}(\bm{R}),\bm{\kappa}'(\bm{R}))
    \tilde{t}_{pd}^*(\bm{\kappa}(\bm{R}'),\bm{\kappa}'(\bm{R}'))
    \nonumber \\
    &\approx
    \left|
    \tilde{t}_{pd}(\bm{\kappa}(\bm{R}),\bm{\kappa}'(\bm{R}))
    \tilde{t}_{pd}^*(\bm{\kappa}(\bm{R}'),\bm{\kappa}'(\bm{R}'))
    \right|,
    \label{eq:appaverage_LSCA}
  \end{align} 
  which assumes that the phases of $\tilde{t}_{pd}(\bm{\kappa}(\bm{R}),\bm{\kappa}'(\bm{R}))$ and $\tilde{t}_{pd}^*(\bm{\kappa}(\bm{R}'),\bm{\kappa}'(\bm{R}'))$ cancel out.
  A detailed discussion on this approximation is made in Appendix~\ref{appendix:local_cs}.

  Making the on-the-energy-shell (on-shell) approximation to $\tilde{t}_{pd}$ using the relative momentum and the energy in the final state (final-state prescription), one obtains 
  \begin{align}
    \left(\frac{\mu_{pd}}{2\pi\hbar^2}\right)^2 
    \left|
    \tilde{t}_{pd}(\bm{\kappa}(\bm{R}), \bm{\kappa}'(\bm{R})) 
    \right|^2
    \approx 
    \frac{d\sigma_{pd}}{d\Omega_{pd}}
    \left(\theta_{pd}(\bm{R}), E_{pd}(\bm{R})\right),
    \label{eq:on-shell}
  \end{align}
  where $\mu_{pd}$ is the reduced energy of the $p$-$d$ system and $d\sigma_{pd}/d\Omega_{pd}$ represents the $p$-$d$ elastic cross section.
  The reduced $T$-matrix with LSCA is obtained as
  \begin{align}
    \bar{T}^{\mathrm{LSCA}}
    &=
    \int d\bm{R}\, 
    \sqrt{\frac{d\sigma_{pd}}{d\Omega_{pd}}(\theta_{pd}(\bm{R}), E_{pd}(\bm{R}))}
    F(\bm{R})\varphi_{d}(\bm{R}).
    \label{eq:T-bar-LSCA}
  \end{align}
  The $p$-$d$ scattering angle $\theta_{pd}(\bm{R})$ and the scattering energy $E_{pd}(\bm{R})$ are determined locally.
  Using $\bar{T}^\mathrm{LSCA}$, the triple differential cross section (TDX) of the ($p$,$pd$) reaction is given by~\cite{Chant77,Wakasa17,Chazono21,Ogata23}
  \begin{align}
    \frac{d^3\sigma^\mathrm{L}_\mathrm{LSCA}}{dE_1^\mathrm{L} d\Omega_1^\mathrm{L} d\Omega_2^\mathrm{L}}
    &=
    \frac{(2\pi)^4}{\hbar v}
    S_d
    F_{\mathrm{kin}}^\mathrm{L}
    \frac{E_1 E_2 E_\mathrm{B}}{E_1^\mathrm{L} E_2^\mathrm{L} E_\mathrm{B}^\mathrm{L}} \frac{1}{2l+1}  \nonumber \\
    &\times \left( \frac{2\pi\hbar^2}{\mu_{pd}} \right) ^2
    \left | \bar{T}^\mathrm{LSCA} \right|^2, 
    \label{eq:tdx-lsca} 
  \end{align}
  \begin{align}
    F_{\mathrm{kin}}^\mathrm{L} 
    \equiv
    \frac{E_1^\mathrm{L}K_1^\mathrm{L}E_2^\mathrm{L}K_2^\mathrm{L}}{(\hbar c)^2} 
    \left[
      1
      +\frac{E_2^{\mathrm{L}}}{E_{\mathrm{B}}^\mathrm{L}}
      +\frac{E_2^\mathrm{L}}{E_\mathrm{B}^\mathrm{L}}
      \frac{\bm{K}_{\mathrm{X}}^{\mathrm{L}} \cdot \bm{K}_{2}^{\mathrm{L}}}
      { \left( K_{2}^{\mathrm{L}} \right)^2 }
    \right]^{-1},
  \end{align}
  with $\bm{K}_\mathrm{X}^\mathrm{L} = \bm{K}_1^\mathrm{L} - \bm{K}_0^\mathrm{L} - \bm{K}_\mathrm{A}^\mathrm{L}$.
  Here, $\Omega_i$ is the solid angle of $\bm{K}_i$ of particle $i$.
  $v$ is the relative velocity between particle 0 and the target A, and $l$ is the relative angular momentum between B and the deuteron in A.
  $S_d$ is the aforementioned deuteron spectroscopic factor.
  As discussed in Sec.~\ref{sec:input}, only the $l=0$ case is discussed in the present work.

  In AMA~\cite{Wakasa17,Yoshida16}, $\bm{K}_i(\bm{R})$ is replaced with its asymptotic momentum $\bm{K}_i$.
  Thus, $\theta_{pd}$, $E_{pd}$, and the $p$-$d$ elementary cross section
  are determined only by the asymptotic momenta as
  \begin{align}
    \left(\frac{\mu_{pd}}{2\pi\hbar^2}\right)^2 
    \left|
    \tilde{t}_{pd}(\bm{\kappa}, \bm{\kappa}') 
    \right|^2
    \approx 
    \frac{d\sigma_{pd}}{d\Omega_{pd}}
    \left(\theta_{pd}, E_{pd}\right).
    \label{eq:pd-cs}
  \end{align}
  In other words, $d\sigma_{pd}/d\Omega_{pd}$ no longer depends on $\bm{R}$. 
  The factorized form of the DWIA TDX is given by~\cite{Chant77,Wakasa17,Chazono21,Ogata23}
  \begin{align}
    \frac{d^3\sigma^\mathrm{L}_\mathrm{AMA}}{dE_1^\mathrm{L} d\Omega_1^\mathrm{L} d\Omega_2^\mathrm{L}}
    &=
    \frac{(2\pi)^4}{\hbar v}
    S_d
    F_{\mathrm{kin}}^\mathrm{L}
    \frac{E_1 E_2 E_\mathrm{B}}{E_1^\mathrm{L} E_2^\mathrm{L} E_\mathrm{B}^\mathrm{L}}
    \frac{1}{2l+1} \nonumber \\
    &\times \left( \frac{2\pi\hbar^2}{\mu_{pd}} \right) ^2
  \frac{d\sigma_{pd}}{d\Omega_{pd}}
    \left | \bar{T}^\mathrm{AMA} \right|^2, 
    \label{eq:tdx-ama}\\
    \bar{T}^\mathrm{AMA}
    &=
    \int d\bm{R}\,
    F(\bm{R})\varphi_{d}(\bm{R}).
    \label{eq:tbar-ama}
  \end{align}
  In this article, we call this the factorized form of the DWIA, not the zero-range DWIA (ZR-DWIA), because the $p$-$d$ elastic cross section $d\sigma_{pd}/d\Omega_{pd}$ results from the $p$-$d$ effective interaction $t_{pd}(\bm{s})$ as seen in Eqs.~(\ref{eq:small-t}) and (\ref{eq:pd-cs}), which is a finite-range interaction.
  It should be noted, however, that as discussed in Sec.~\ref{sec:input}, we directly use experimental data of the $p$-$d$ elastic cross section for $d\sigma_{pd}/d\Omega_{pd}$ in Eqs.~(\ref{eq:T-bar-LSCA}) and (\ref{eq:tdx-ama}).
  Therefore, an explicit form of $t_{pd}(\bm{s})$ is not considered in the present study. 

\section{Result and discussion}
  \label{sec:result}
  \subsection{Numerical input} 
    \label{sec:input}
\begin{figure*}[htpb]
      \centering
      \includegraphics[width=\textwidth]{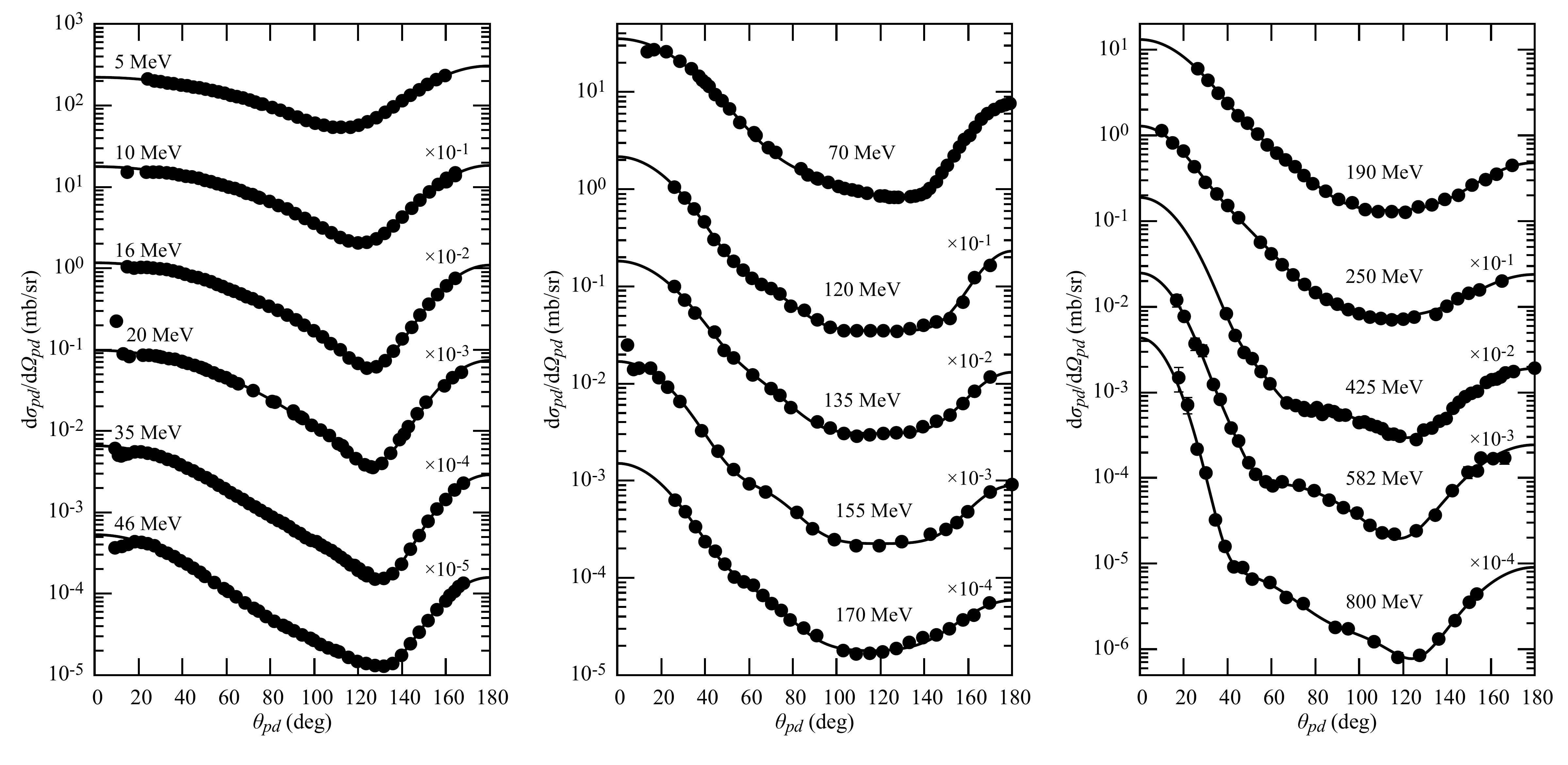}  
      \caption{Angular distribution of the $p$-$d$ elastic scattering cross section.
        The horizontal axis is the center-of-mass scattering angle of the $p$-$d$ system.
        The experimental data taken from Refs.~\cite{KKuroda64, SNBunker68, TACahill71, NEBooth71, ETBoschitz72, EWinkelmann80, KSagara94, KSekiguchi02, KHatanaka02, KErmisch05} are represented by filled circles.
        The solid lines show the results of the fitting.
        See Eq.~(\ref{eq:pd_fit}) and Table~\ref{tab:pd_fit_param} for details.
        }
      \label{fig:pd_fit}
    \end{figure*}

    The differential cross section of the $p$-$d$ elastic scattering is prepared numerically by fitting the experimental data from $5$ to $800$~MeV with respect to the scattering energy and the angle as shown in Fig.~\ref{fig:pd_fit}.
    The experimental data of the $p$-$d$ elastic scattering are taken from Refs.~\cite{KKuroda64, SNBunker68, TACahill71, NEBooth71, ETBoschitz72, EWinkelmann80, KSagara94, KSekiguchi02, KHatanaka02, KErmisch05}.
    The function used in the fitting is
    \begin{align}
      \frac{d\sigma_{pd}}{d\Omega_{pd}}
      =
      \sum_{j=1}^{j_\textrm{max}} a_j \exp \left[ - \left( \frac{\theta_{pd} - c_j}{b_j} \right)^2 \right] + d_0,
      \label{eq:pd_fit}
    \end{align}
    where $j_\textrm{max}$ is $2$ for $5$--$20$~MeV, $3$ for $35$--$250$~MeV, and $4$ for $425$--$800$~MeV.
    All the parameters $a_j$, $b_j$, $c_j$, and $d_0$ are listed in Table~\ref{tab:pd_fit_param}.
    It should be noted that $c_1$ ($c_{j_\textrm{max}}$) is fixed at $0^\circ$ ($180^\circ$).
    \begin{table}[ht!]
      \centering
      \begin{tabular}{rrrrrrrrrrr}
        \hline\hline
        $E_{pd}$ && $j$ && \multicolumn{1}{c}{$a_j$} && \multicolumn{1}{c}{$b_j$} && \multicolumn{1}{c}{$c_j$} && \multicolumn{1}{c}{$d_0$} \\
        \hline
          5 &&  1  && 222.1560 && 86.8088 &&   0.0000 && 0.0000 \\
            &&  2  && 302.8700 && 37.5459 && 180.0000 && \\
         10 &&  1  && 177.2360 && 78.7301 &&   0.0000 && 0.0000 \\
            &&  2  && 182.1080 && 30.6876 && 180.0000 && \\
         16 &&  1  && 116.5830 && 70.8749 &&   0.0000 && 0.0000 \\
            &&  2  && 108.7870 && 26.0295 && 180.0000 && \\
         20 &&  1  &&  97.9835 && 67.8439 &&   0.0000 && 0.0000 \\
            &&  2  &&  73.3261 && 25.8404 && 180.0000 && \\
         35 &&  1  &&  65.6339 && 52.7178 &&   0.0000 && 0.0000 \\
            &&  2  &&   2.7094 && 35.3458 &&   93.5447 && \\
            &&  3  &&  28.8682 && 24.2590 && 180.0000 && \\
         46 &&  1  &&  52.7869 && 44.5933 &&   0.0000 && 0.0000 \\
            &&  2  &&   2.7551 && 48.1972 &&  80.0000 && \\
            &&  3  &&  15.7479 && 24.9981 && 180.0000 && \\
        \hline
         70 &&  1  &&  34.6225 && 38.2736 &&   0.0000 && 0.7461 \\
            &&  2  &&   0.5250 && 29.8396 &&  80.0000 && \\
            &&  3  &&   6.9138 && 21.1479 && 180.0000 && \\
        120 &&  1  &&  21.2430 && 30.7048 &&   0.0000 && 0.3485 \\
            &&  2  &&   0.4932 && 19.8301 &&  64.7798 && \\
            &&  3  &&   1.9542 && 18.4435 && 180.0000 && \\
        135 &&  1  &&  17.8934 && 31.7476 &&   0.0000 && 0.3064 \\
            &&  2  &&   0.5316 && 20.8098 &&  61.0116 && \\
            &&  3  &&   0.9988 && 21.4401 && 180.0000 && \\
        155 &&  1  &&  16.6284 && 29.5589 &&   0.0000 && 0.2246 \\
            &&  2  &&   0.4812 && 20.5405 &&  63.6499 && \\
            &&  3  &&   0.6649 && 21.1364 && 180.0000 && \\
        170 &&  1  &&  14.8897 && 27.9947 &&   0.0000 && 0.1761 \\
            &&  2  &&   0.5273 && 23.9095 &&  55.2300 && \\
            &&  3  &&   0.4100 && 26.4076 && 180.0000 && \\
        \hline
        190 &&  1  &&  13.0640 && 28.5562 &&   0.0000 && 0.1305 \\
            &&  2  &&   0.5110 && 26.0810 &&  50.0000 && \\
            &&  3  &&   0.3499 && 28.3901 && 180.0000 && \\
        250 &&  1  &&   9.6172 && 20.2655 &&   0.0000 && 0.0741 \\
            &&  2  &&   3.1887 && 40.5310 &&   0.0000 && \\
            &&  3  &&   0.1642 && 30.3782 && 180.0000 && \\
        425 &&  1  &&  18.8803 && 21.7724 &&   0.0000 && 0.1305 \\
            &&  2  &&   0.1294 && 13.0634 &&  49.2562 && \\
            &&  3  &&   0.0613 && 38.2106 &&  80.0000 && \\
            &&  4  &&   0.1900 && 34.5092 && 180.0000 && \\
        582 &&  1  &&  24.7257 && 18.9076 &&   0.0000 && 0.1305 \\
            &&  2  &&   0.1519 && 10.1242 &&  38.0159 && \\
            &&  3  &&   0.0858 && 41.6712 &&  60.0000 && \\
            &&  4  &&   0.2437 && 33.0826 && 180.0000 && \\
        800 &&  1  &&  43.1368 && 15.4699 &&   0.0000 && 0.1305 \\
            &&  2  &&   0.0780 && 33.9306 &&  40.0000 && \\
            &&  3  &&   0.0111 && 25.3878 && 100.0000 && \\
            &&  4  &&   0.0904 && 30.4011 && 180.0000 && \\
        \hline\hline
      \end{tabular}
      \caption{Numerical value of the parameters in Eq.~(\ref{eq:pd_fit}). $E_{pd}$ is given in MeV; $a_j$ and $d_0$ are in mb/sr; $\theta_{pd}$, $b_j$, and $c_j$ are in degrees.}
      \label{tab:pd_fit_param}
    \end{table}
    In the present calculation, the fitted $p$-$d$ elastic cross section is interpolated to the required energy and scattering angle.
    Compared to the previous analysis~\cite{Samanta82} in which only the $p$-$d$ scattering data at $22$, $35$, and $46$~MeV~\cite{SNBunker68} are used, the wide coverage of the $p$-$d$ scattering energies improves the reliability of the elementary cross section of the ($p,pd$) reaction at $E_{pd} > 46~\mathrm{MeV}$. 

    The proton distorted waves, $\chi_{0}$ and $\chi_{1}$, are obtained as scattering waves under the phenomenological optical potential by Koning and Delaroche~\cite{Koning03}.
    For the deuteron distorted wave $\chi_{2}$, the optical potential proposed by An and Cai~\cite{An06} is adopted.
    The nonlocality correction to the distorted waves is taken into account by the Perey factor~\cite{Perey62} with the range parameter $\beta = 0.85$~fm for the proton and $\beta = 0.54$~fm for the deuteron~\cite{TWOFNR}.

    $\varphi_{d}$ is obtained as a bound state wave function under the Woods-Saxon shaped potential with the range parameter $R_0 = 1.41\times14^{1/3}$~fm and the diffuseness parameter $a_0 = 0.65$~fm~\cite{Samanta82}.
    The depth parameter of the potential is adjusted to reproduce the effective deuteron separation energy of $24.7$~MeV, which is the sum of the deuteron separation energy of $^{16}$O ($20.74$~MeV) and the excitation energy of the $^{14}$N residue ($3.95$~MeV).
    As discussed in Ref.~\cite{Samanta82}, the $^{16}$O($p$,$pd$)$^{14}$N$_{1^{+}}$ data show a typical $s$-wave peak at the recoilless condition, in which the reaction residue is at rest in the final state.
    Following Ref.~\cite{Samanta82}, we only consider $l=0$ for the $d$-$^{14}$N$_{1^{+}}$ relative angular momentum.

  \subsection{Refraction effect in $^{16}\mathrm{O}(p,pd)^{14}\mathrm{N}$}
    \label{sec:refraction}
\begin{figure}[htpb]
      \centering
      \includegraphics[width=0.45\textwidth]{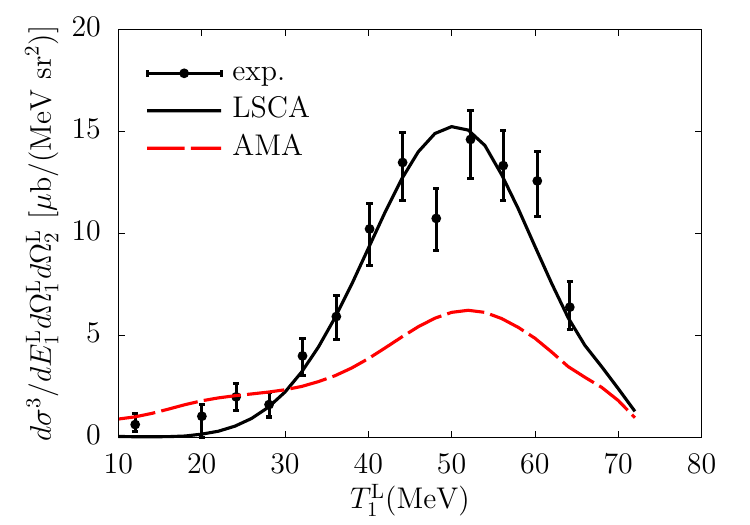}
      \caption{TDXs of the $^{16}$O($p$,$pd$)$^{14}$N$_{1^{+}}$ reaction calculated with LSCA and AMA in comparison with the experimental data~\cite{Samanta82}. 
        Both are multiplied by $S_d = 1.74$.}
      \label{fig:100mev}
    \end{figure}

    The TDXs of the $^{16}$O($p$,$pd$)$^{14}$N$_{1^{+}}$ at $101.3~\mathrm{MeV}$ calculated with LSCA and AMA, and the experimental data~\cite{Samanta82} are shown in Fig.~\ref{fig:100mev} as a function of the proton emission energy $T_1^\mathrm{L}$.
    The proton and deuteron emission angles are $\theta_1^\mathrm{L} = 40.1^\circ$ and $\theta_2^\mathrm{L} = 40.0^\circ$, respectively, and the kinematics are in coplanar.
    The TDX with LSCA clearly shows better agreement with the data than that with AMA in their shape, showing that the refraction effect is important.
    The experimental spectroscopic factor $S_d = 1.74$ is obtained by fitting the theoretical curve of LSCA to the data.
    The TDX using AMA is also multiplied by $S_d = 1.74$ for comparison.
    The obtained $S_d$ is consistent with the 1$p$ shell model prediction at early ages~\cite{Cohen70} $S_d = 1.75$, which is referred to in the former experimental analyses~\cite{Samanta82,Grossiord77}, and is comparable with 1.43 determined in the ZR-DWIA analysis~\cite{Joshi16}.

    As discussed in Sec.~\ref{sec:framework}, LSCA should partially incorporate the finite-range nature of the $p$-$d$ effective interaction into the $p$-$d$ cross section through the $\bm{R}$ dependence of $\bm{\kappa}(\bm{R})$ and $\bm{\kappa}'(\bm{R})$.
    As seen in the difference between the LSCA and AMA results in Fig.~\ref{fig:100mev}, the shape of the distribution strongly depend on how $\theta_{pd}$ and $E_{pd}$ are evaluated.
    The present analysis shows that LSCA is valid for reproducing both the shape and the height of the experimental data and the cross section with a reasonable $S_d$ as long as $d\sigma_{pd}/d\Omega_{pd}$ is prepared precisely as in Fig.~\ref{fig:pd_fit}.

    The origin of the inconsistency between the present AMA and the ZR-DWIA in Ref.~\cite{Joshi16} is not clear. 
    In the present AMA calculation, the shape of the TDX cannot be reproduced as shown in Fig.~\ref{fig:100mev}, while the ZR-DWIA calculation gives good agreement with the data as shown in Fig.~2 of Ref.~\cite{Joshi16}.
    Although the detail of the ZR-DWIA calculation is not described in Ref.~\cite{Joshi16}, the inconsistency may arise from the different treatment of the $p$-$d$ elementary process.
    In the present study we use the fitted function of the experimental $p$-$d$ differential cross sections as shown in Fig.~\ref{fig:pd_fit}, while in Ref.~\cite{Joshi16} they constructed $t_{pd}(\bm{s})$ [denoted as $t_L(E,r)$ in the reference] based on the $p$-$d$ optical potential~\cite{Kim64,Hinterberger68,SNBunker68}.
    Also the difference between the present LSCA result and the FR-DWIA in Ref.~\cite{Joshi16} is not clear.
    In two types of the FR-DWIA analysis in Ref.~\cite{Joshi16}, all attractive and repulsive core, $S_d = 0.47$ and $0.12$ are obtained, respectively, which are both smaller than the present value of 1.74, suggesting again that the difference in the treatment of the $p$-$d$ elementary process.
    It should be noted here that the diffraction effect should be naturally implemented in FR-DWIA by the explicit distance-dependent $p$-$d$ $t$ matrix and the six-dimensional integral. 
    The AMA formalism of the present work is essentially identical to that used in the analysis of the original experimental data~\cite{Samanta82}.
    The inconsistency between the present AMA result and the curve in Fig.~2(a) of Ref.~\cite{Samanta82} is found to be due to the differences in the optical potentials and the $p$-$d$ cross section used as inputs. Details are discussed in Appendix~\ref{appendix:Samanta}.

\begin{figure}[htpb]
      \centering
      \includegraphics[width=0.45\textwidth]{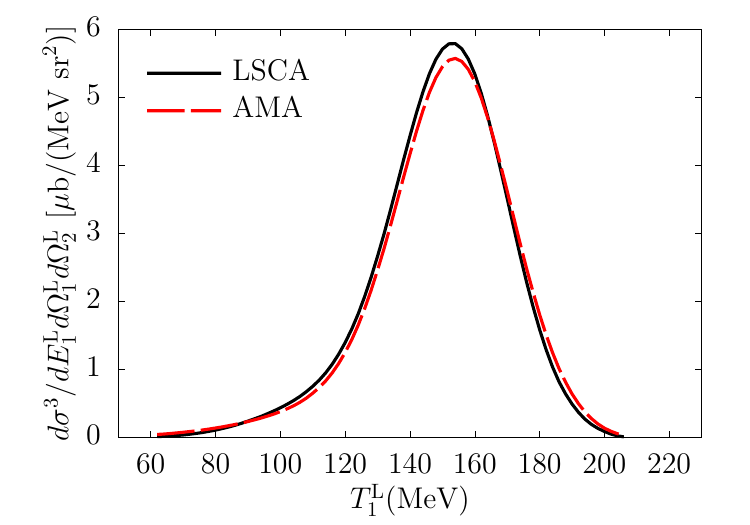}
      \caption{Same as Fig.~\ref{fig:100mev} but at 250 MeV.}
      \label{fig:250mev}
    \end{figure}

    To investigate the impact of the deuteron refraction at higher energies, we consider the same reaction at 250~MeV; the results are shown in Fig.~\ref{fig:250mev}.
    The kinematics conditions are fixed at $\theta_1^\mathrm{L} = 47.1^\circ$ and $\theta_2^\mathrm{L} = 48.8^\circ$, which gives the recoilless condition at around $T_1^\mathrm{L} = 150$~MeV.
    Since there is no experimental data at higher energies, only the theoretical curves are shown assuming $S_d = 1$ in Fig.~\ref{fig:250mev}.
    It is clearly shown that the result using LSCA is almost equivalent to that using AMA at 250~MeV, in contrast to the 101.3~MeV case in Fig.~\ref{fig:100mev}.
    This result is consistent with the previous work on the $^{120}$Sn($p$,$p\alpha$)$^{116}$Cd reaction at 392~MeV~\cite{Yoshida16}.

\section{Summary}
  \label{sec:summary}
  The refraction effect on the elementary process of the $(p,pd)$ reaction has been investigated by using LSCA in the DWIA framework.
  It is found that the effect gives a significant change in the shape of the energy sharing distribution of the $^{16}$O($p$,$pd$)$^{14}$N reaction at 101.3~MeV.
  The present analysis using LSCA gives $S_d = 1.74$, which is consistent with 1$p$ shell model prediction 1.75~\cite{Cohen70} referred in Refs.~\cite{Samanta82,Grossiord77}, and comparable with 1.43 by the previous ZR-DWIA analysis~\cite{Joshi16}.
  On the other hand, at 250~MeV, the results with LSCA and AMA are similar, showing that the local $p$-$d$ kinematics has little effect on the TDX and the commonly used factorized form of the DWIA with AMA can be safely applied to the reaction analysis at this energy.

  As discussed in Refs.~\cite{Chazono21,Chazono22}, the description of the spatial distribution and the radial magnitude of $\varphi_{d}(\bm{R})$ based on microscopic structure theories is still challenging but important for the quantitative discussion of the deuteron-like $p$-$n$ pair in nuclei.
  A combination of the present work and CDCCIA~\cite{Chazono22}, which takes into account the breakup and the reformation process of the $p$-$n$ pair, will be important for the accurate description of the $(p,pd)$ reaction. 
  The development of such framework as well as a numerical code is ongoing.

\begin{acknowledgments}
  We thank T.~Uesaka and S.~Ogawa for fruitful discussions. This work is supported in part by Grants-in-Aid for Scientific Research from the JSPS (Grants No. JP20K14475, No. JP21H00125, and No. JP21H04975).
\end{acknowledgments}


\appendix
\section{Local cross section approximation}
\label{appendix:local_cs}
Equation~(\ref{eq:appaverage_LSCA}) assumes the cancellation of the complex phase factors of $\tilde{t}_{pd}(\bm{\kappa}(\bm{R}),\bm{\kappa}'(\bm{R}))$ and $\tilde{t}_{pd}^*(\bm{\kappa}(\bm{R}'),\bm{\kappa}'(\bm{R}'))$.
In this appendix we consider this approximation in terms of $t_{pd}(\bm{s})$, the effective interaction in the coordinate space.
Here, we assume that $t_{pd}(\bm{s})$ is the $p$-$d$ effective interaction parameterized so as to reproduce the $p$-$d$ elastic cross section only with the central term:
\begin{align} 
  t_{pd}(E,\bm{s}) 
  &=
  \left[ v(E) + iw(E) \right] f(s),
  \label{eq:t-parameterized}
\end{align}
with $E$ being the incident energy, and $v(E)$ and $w(E)$ are the energy dependent real and imaginary parts of the central term strength.
To show the energy dependence of $t_{pd}$ clearly, $E$ is added to its argument.
The radial dependence of the effective interaction is expressed by the real function $f(s)$.
The Fourier transform of Eq.~(\ref{eq:t-parameterized}) is given by
\begin{align}
  \tilde{t}_{pd}(E,q) 
  &=
  \left[ v(E) + iw(E) \right] \tilde{f}(q), 
  \label{eq:t_pd-momentum-space}\\
  \tilde{f}(q) 
  &\equiv
  4\pi \int j_0(qs)f(s)s^2\,ds,
\end{align}
where $\bm{q}$ is the conjugate momentum of the coordinate $\bm{s}$, and $j_0$ is the zeroth spherical Bessel function.

Using Eq.~(\ref{eq:t_pd-momentum-space}) with $\bm{q}(\bm{R}) \equiv \bm{\kappa}(\bm{R}) - \bm{\kappa}'(\bm{R})$, the left- and the right-hand sides of Eq.~(\ref{eq:appaverage_LSCA}) are rewritten as
\begin{align} 
  &\tilde{t}_{pd}(\bm{\kappa}(\bm{R}),\bm{\kappa}'(\bm{R}))
  \tilde{t}_{pd}^*(\bm{\kappa}(\bm{R}'),\bm{\kappa}'(\bm{R}'))
  \nonumber \\
  &\quad=\tilde{t}_{pd}(E,\bm{q}(\bm{R}))\tilde{t}_{pd}^*(E',\bm{q}(\bm{R}')) \nonumber \\
  &\quad=
  \left[ v(E) + iw(E) \right] \left[ v(E') - iw(E') \right]  \nonumber  \\
  &\qquad\,\times \tilde{f}(q(\bm{R})) \tilde{f}(q(\bm{R}')) 
\end{align} 
and
\begin{align} 
  &\left|
  \tilde{t}_{pd}(\bm{\kappa}(\bm{R}),\bm{\kappa}'(\bm{R}))
  \tilde{t}_{pd}^*(\bm{\kappa}(\bm{R}'),\bm{\kappa}'(\bm{R}'))
  \right|
  \nonumber \\
  &\quad=
  \sqrt{ v^2(E) + w^2(E) } \sqrt{ v^2(E') + w^2(E') } \nonumber \\
  &\qquad\times\tilde{f}(q(\bm{R})) \tilde{f}(q(\bm{R}')),
\end{align} 
respectively.
Thus, the approximation of Eq.~(\ref{eq:appaverage_LSCA}) implies
\begin{align} 
  v(E) + iw(E) \approx v(E') + iw(E'),
  \label{eq:v_and_w_approx}
\end{align} 
that the energy dependence of $\tilde{t}_{pd}$ through $v(E)$ and $w(E)$ is negligible for different $\bm{R}$, compared to the $q(\bm{R})$ dependence.
It should be noted that no approximation is made to $\tilde{f}(q(\bm{R}))$ or $\tilde{f}(q(\bm{R}'))$.
Therefore, the angular dependence of $\tilde{t}_{pd}$ is taken into account properly in this approximation.
In the present study, this approximation is expected to be good since the energy dependence of the $p$-$d$ elastic cross section is weak compared to its angular dependence, as shown in Fig.~\ref{fig:pd_fit}.

\section{On discrepancy between Samanta \textit{et al.} and the present analysis}
\label{appendix:Samanta}
The theoretical frameworks of the present AMA result (see Fig.~\ref{fig:100mev} of the present work) and the curve in Fig.~2 (a) of Ref.~\cite{Samanta82} are essentially the same, but a clear discrepancy can be seen in these results, in both shape of the TDX and the extracted $S_d = 1.74$ and $0.43$.
In this appendix, we show that the discrepancy is due to the differences in the $p$-$d$ elastic cross section and the optical potentials used as the inputs in these analyses.

In the previous analysis in Ref.~\cite{Samanta82}, the $p$-$d$ elastic scattering cross section data at 22, 35, and 46~MeV~\cite{SNBunker68} were used, as mentioned in Sec.~\ref{sec:input} of the present paper.
By the AMA analysis of the present work, it was found that in the $^{16}$O($p$,$pd$)$^{14}$N kinematics of Fig.~\ref{fig:100mev} of the present work and Fig.~2 (a) of Ref.~\cite{Samanta82}, the corresponding $p$-$d$ scattering energy lies in the range $35 \lesssim E_{pd} \lesssim 70$~MeV, which increases monotonically with increasing $T_1^\mathrm{L}$.
At $T_1^\mathrm{L} \approx 35$~MeV, $E_{pd}$ exceeds 46~MeV and thus the $p$-$d$ cross section is extrapolated in $T_1^\mathrm{L} \gtrsim 35$~MeV, if the $p$-$d$ cross section data at 22, 35, and 46~MeV are used.

For the optical potentials, in the previous analysis of Ref.~\cite{Samanta82}, the parameter sets determined from the $p$-$^{12}$C scattering at 100~MeV~\cite{Li68} and 75~MeV~\cite{Rolland66} are adopted for the initial $p$-$^{16}$O and the final $p$-$^{14}$N distorted waves, respectively.
For the $d$-$^{14}$N in the final channel, the $d$-$^{14}$N optical potential parameters determined at 28~MeV~\cite{Gaillard68} are used.
Due to the difference in the available data and the optical potential parameters, we found that the result differs from that obtained with the modern input in the present study.

\begin{figure}[htpb]
    \centering
    \includegraphics[width=0.45\textwidth]{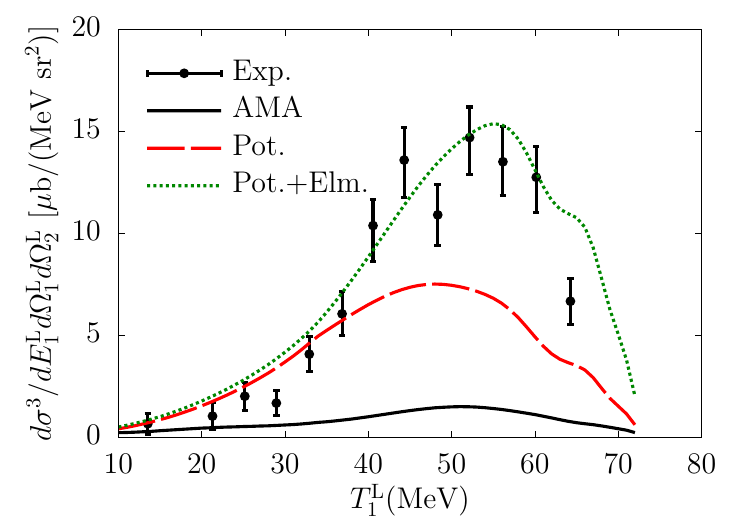}
    \caption{
      Comparison of AMA TDXs of the present work and that using the inputs and $S_d = 0.43$ of Samanta \textit{et al}.~\cite{Samanta82}. 
      The sold line is the AMA result of the present work (same as the dashed line in Fig.~\ref{fig:100mev} but with $S_d = 0.43$), the dashed line shows the same calculation but using the optical potentials used Samanta \textit{et al}.
      The dotted line shows the AMA calculations using both the optical potentials and the $p$-$d$ cross section used in Samanta \textit{et al}.
    }
    \label{fig:input_comp}
  \end{figure}

In Fig.~\ref{fig:input_comp}, we show the AMA results calculated with the optical potentials and the $p$-$d$ cross section used in the earlier work by Samanta \textit{et al.}~\cite{Samanta82} as mentioned above.
Note that $S_d = 0.43$ from Samanta \textit{et al}.~\cite{Samanta82} is applied to all curves in Fig.~\ref{fig:input_comp} for comparison.
The dashed line in Fig.~\ref{fig:input_comp} shows the AMA result using the optical potentials used in Samanta \textit{et al}.~\cite{Samanta82}.
One sees a remarkable increase of the TDX compared to the present AMA result (solid line), while the shape of the curve remains similar.
It is found that the changes in the $p$-$^{16}$O, $p$-$^{14}$N, and $d$-$^{14}$N potentials cause a change in the height of the TDX by factors of 1.2, 2.0, and 2.2, respectively.
Furthermore, using the $p$-$d$ cross section which is adopted in Ref.~\cite{Samanta82}, we obtain the TDX as shown by the dotted line in Fig.~\ref{fig:input_comp}.
It can also be seen that the difference in the $p$-$d$ cross section changes both the magnitude and the shape of the TDX.
In particular, the shape changes in $T_1^\mathrm{L} \gtrsim 35$~MeV, indicating that the extrapolation beyond $E_{pd}=46$~MeV makes the difference. 
The agreement in shape with the experimental data around the peak seems rather coincidental, since the energy and angular dependence of the $p$-$d$ cross section should be better described in the present work.
The dotted line in Fig.~\ref{fig:input_comp} of the present work is almost identical to Fig.~2 (a) of Ref.~\cite{Samanta82}, and it can be concluded that the difference between the present AMA result and the DWIA calculation in Ref.~\cite{Samanta82} originates from the differences of the optical potentials and the $p$-$d$ cross section used as the inputs.

\bibliography{./bib/ref_normal}

\end{document}